\documentclass[Physsubmission, Phys]{SciPost}

\hypersetup{
    colorlinks,
    linkcolor={red!50!black},
    citecolor={blue!50!black},
    urlcolor={blue!80!black}
}
\usepackage{braket}
\usepackage[bitstream-charter]{mathdesign}
\DeclareSymbolFont{usualmathcal}{OMS}{cmsy}{m}{n}
\DeclareSymbolFontAlphabet{\mathcal}{usualmathcal}

\begin{document}

\begin{center}{\Large \textbf{
Valence quark distributions of light mesons in light-cone quark model
}}\end{center}

\begin{center}
Satvir Kaur\textsuperscript{1},
Narinder Kumar\textsuperscript{2},
Jiangshan Lan\textsuperscript{3,4,5},
Chandan Mondal\textsuperscript{3,4} and
Harleen Dahiya\textsuperscript{1$\star$}
\end{center}

\begin{center}
{\bf 1} Department of Physics, Dr. B. R. Ambedkar National Institute of Technology, Jalandhar 144011, India
\\
{\bf 2} Department of Physics, Doaba College, Jalandhar 144004, India
\\
{\bf 3} Institute of Modern Physics, Chinese Academy of Sciences, Lanzhou 730000, China
\\
{\bf 4} School of Nuclear Science and Technology, University of Chinese Academy of Sciences, Beijing 100049, China
\\
{\bf 5} Lanzhou University, Lanzhou 730000, China
\\
* dahiyah@nitj.ac.in
\end{center}

\begin{center}
\today
\end{center}


\definecolor{palegray}{gray}{0.95}
\begin{center}
\colorbox{palegray}{
  \begin{tabular}{rr}
  \begin{minipage}{0.1\textwidth}
    \includegraphics[width=22mm]{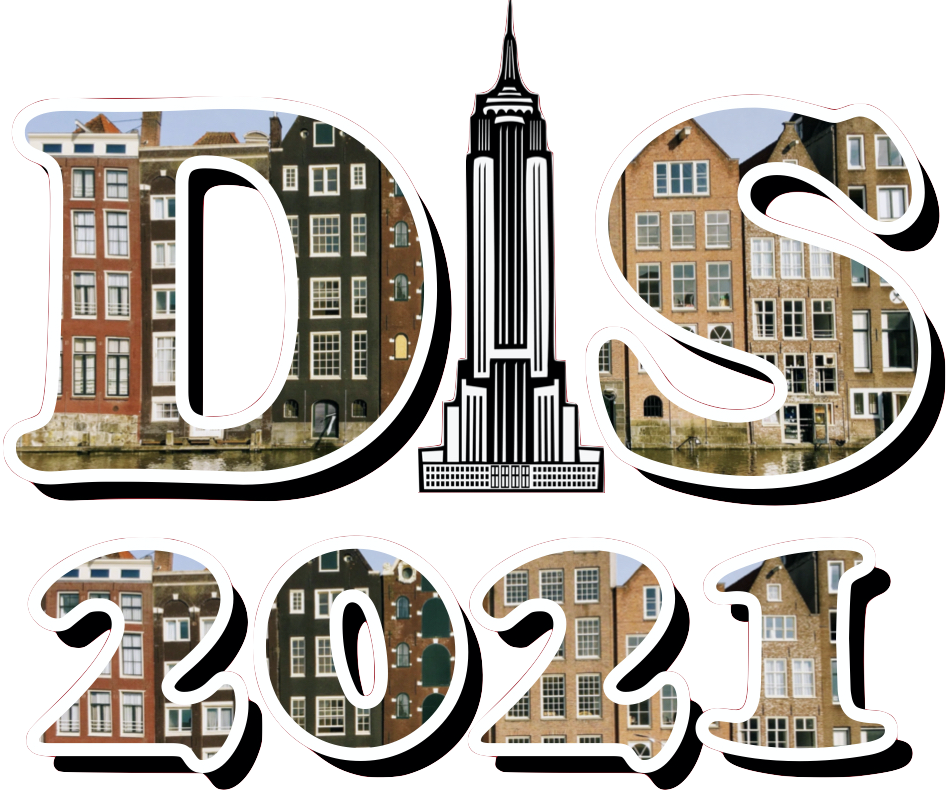}
  \end{minipage}
  &
  \begin{minipage}{0.75\textwidth}
    \begin{center}
    {\it Proceedings for the XXVIII International Workshop\\ on Deep-Inelastic Scattering and
Related Subjects,}\\
    {\it Stony Brook University, New York, USA, 12-16 April 2021} \\
    \doi{10.21468/SciPostPhysProc.?}\\
    \end{center}
  \end{minipage}
\end{tabular}
}
\end{center}

\section*{Abstract}
{\bf
In order to investigate the tomographical structure of light pseudoscalar mesons, particularly, pion and kaon, we study
the valence quark distribution function (PDF) and the generalized parton distributions (GPDs) using light-cone quark model (LCQM).
}

%

\section{Introduction}
\label{sec:intro}
The parton distribution functions (PDFs) \cite{soper, martin} which encode the distribution with respect to the parton's longitudinal momentum and its polarization; and the generalized parton distributions (GPDs) \cite{diehl, garcon, geoke, belitsky} give the three-dimensional (3D) spatial distribution of the partons in the transverse direction of the hadron motion. The PDFs being the function of parton's longitudinal momentum fraction ($x$) provide the one-dimensional information on the hadron structure and the GPDs, which are three-dimensional distributions, are function of $x$, the longitudinal momentum transferred ($\zeta$) and the total momentum transferred from hadron's initial state to final state ($t$).

Experimentally, GPDs are accessible through the deeply virtual Compton scattering (DVCS) \cite{dvcs1, dvcs2, dvcs3} and deeply virtual meson production (DVMP) \cite{dvmp1, dvmp2}. GPDs are also possible to extract via the $\rho$-meson photoproduction \cite{meson-photo-prod2}, time-like Compton scattering \cite{time-like1, time-like3}, exclusive pion or photon-induced lepton pair-production \cite{lp1, gpd-exp}.   Theoretically, the pion DAs and PDFs have been studied using various approaches \cite{cz, gk, pdf-njl, da-constituent, pdf-chiral},  phenomenological models \cite{Frederico:1994dx,Shigetani:1993dx,Weigel:1999pc}, anti-de Sitter (AdS)/QCD models \cite{zero-skewness,Ahmady:2018,deTeramond:2018ecg} and lattice QCD~\cite{Brommel:2006zz,Detmold:2003tm,Oehm:2018jvm,Lin:2017snn,Joo:2019bzr}. 
The pion GPDs have been computed \cite{covariant-n-constituent, gpd-chiral}. Recently, the pion GPDs for non-zero skewedness \cite{non-zero-skewness} and for zero skewedness \cite{zero-skewness} have been studied with AdS/QCD approach.

In the present work, we have used the light-front spin-improved wave functions of light pseudoscalar mesons to investigate the 
the PDFs for the case of pion and kaon. Also, we have evaluated the chirally-even GPD $H(x, \zeta, t)$ describing the distribution of an unpolarized quark and chirally-odd GPDs $E_T(x, \zeta, t)$ corresponding to the distribution of a transversely polarized quark inside the pseudoscalar meson from the overlap of light-front wave functions (LFWFs). 
The LFWFs in the light-cone quark model (LCQM) are obtained via adopting the Brodsky-Huang-Lepage (BHL) prescription. 
Further, the spin-improved wave functions are derived using the Melosh-Wigner transformation.
 The LCQM is successful in explaining the electromagnetic form factors of pion and kaon and it can be enhanced to study other distributions of the quarks in the mesons.

\section{Light-Cone Quark Model (LCQM)}
The hadron eigenstate is connected with the multi-particle Fock eigenstates $|n \rangle$ 
containing $n$ constituents where the $i^{th}$ constituent holds the longitudinal momentum fraction $x_i=\frac{k_i^+}{P^+}$, the transverse momentum $\textbf{k}_{\perp i}$ and  helicity $\lambda_i$ \cite{meson-state}. It is normalized as $\langle{n;{k'}_i^+, {\textbf{k}'}_{\perp i},\lambda'_i} | {n;{k}_i^+, {\textbf{k}}_{\perp i},\lambda_i}\rangle$ $=\prod_{i=1}^n 16 \pi^3 k_i^+ \delta({k'}_i^+ -k_i^+) \delta^{(2)} ({\textbf{k}'}_{\perp i}-{\textbf{k}'}_{\perp i})\delta_{\lambda'_i \lambda_i}.$
 The complete wave function in LCQM can be expressed as $\psi_{S_z}^F(x, \textbf{k}_\perp, \lambda_1, \lambda_2) = \varphi(x, \textbf{k}_\perp)  \chi_{S_z}^F(x,\textbf{k}_\perp, \lambda_1, \lambda_2)$, where $\varphi$ and $\chi$ are respectively the momentum space and spin wave functions. The superscript $F$ denotes the front form. The light-cone spin wave function of pseudoscalar `$\mathcal{P}$' meson (pion or kaon) is defined as
\begin{eqnarray}
	\chi^\mathcal{P}(x,\textbf{k}_\perp)=\sum_{\lambda_1, \lambda_2}\kappa_{S_z}^F(x,\textbf{k}_\perp,\lambda_1, \lambda_2) \chi_1^{\lambda_1}(F) \chi_2^{\lambda_2}(F),
\end{eqnarray}
where $S_z$ and $\lambda$ are the meson's spin projection and quark helicity, respectively. The momentum space wave functions for pion and kaon are respectively expressed as 
\begin{eqnarray}
	\varphi^\pi(x,\textbf{k}_\perp)&=&A^\pi \ {\rm exp}\bigg[-\frac{1}{8 \beta_\pi^2} \frac{{\bf k}^2_\perp+m^2} {x(1-x)} \bigg], \label{bhl-pi} \\
	\varphi^K(x,\textbf{k}_\perp)&=& A^K \ {\rm exp} \Bigg[-\frac{\frac{\textbf{k}^2_\perp+m_1^2}{x}+\frac{\textbf{k}^2_\perp+m_2^2}{1-x}}{8 \beta_K^2} 
	-\frac{(m_1^2-m_2^2)^2}{8 \beta_K^2 \bigg(\frac{\textbf{k}^2_\perp+m_1^2}{x}+\frac{\textbf{k}^2_\perp+m_2^2}{1-x}\bigg)}\Bigg],
	\label{bhl-k}
\end{eqnarray}
where $A^\pi$ and $A^K$ define the normalization constants in case of pion and kaon.

\section{Parton distribution functions (PDFs)}
\begin{figure}
\centering
\begin{minipage}[c]{1\textwidth}
(a)\includegraphics[width=.45\textwidth]{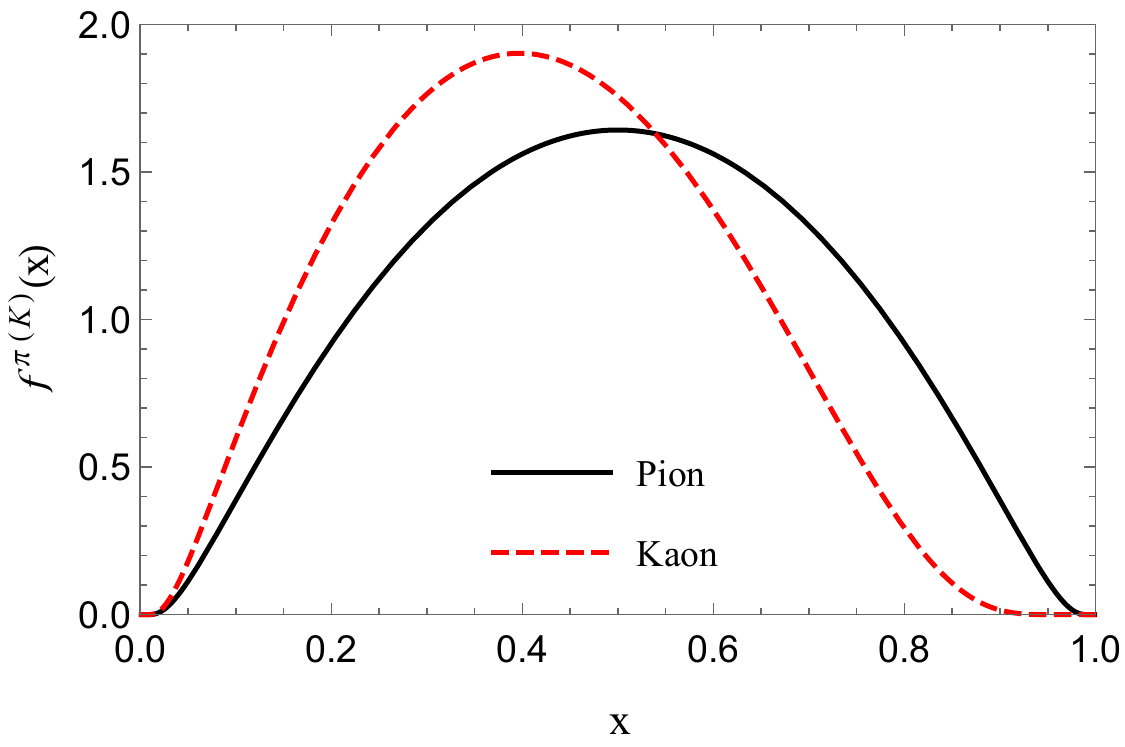}
(b)\includegraphics[width=.45\textwidth]{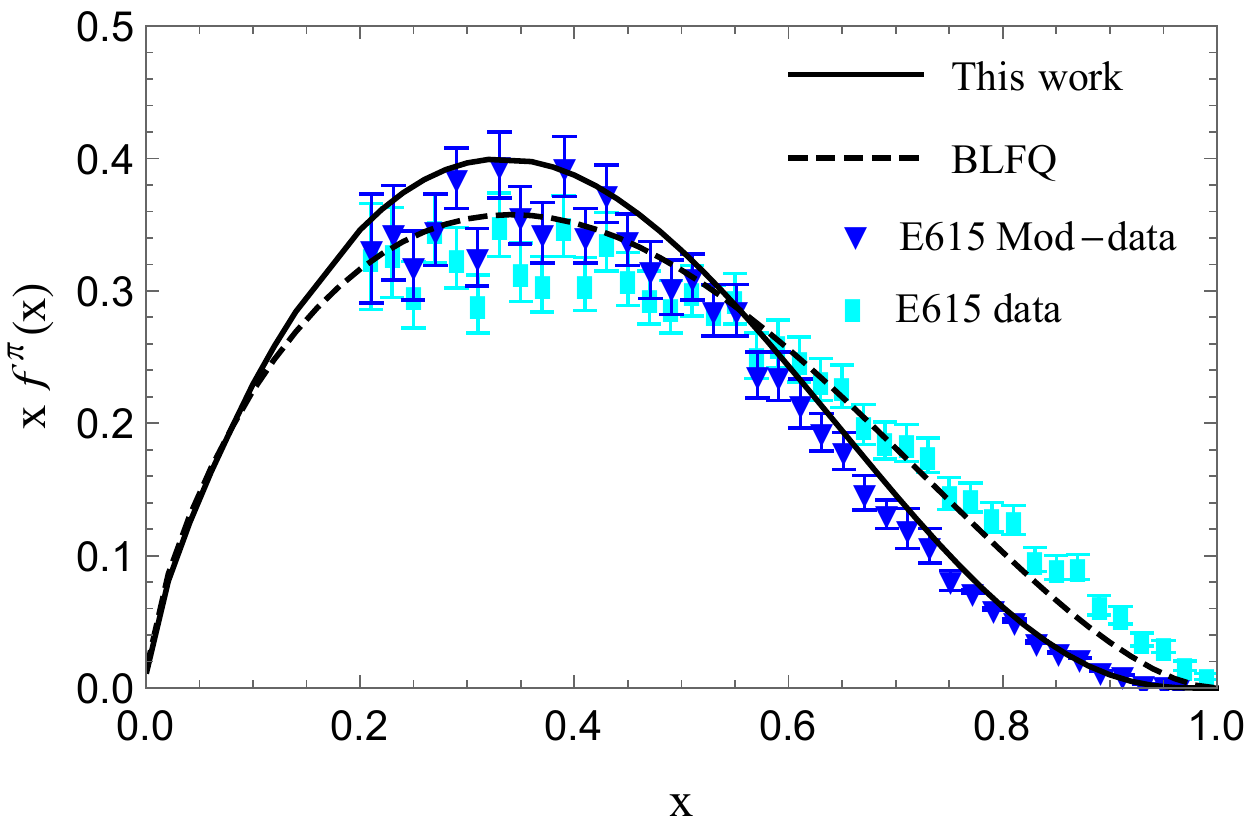}
\end{minipage}
\caption{(a) The unpolarized pion and kaon PDFs at model scale. (b) The unpolarized PDF in LCQM evolved from the initial scale $\mu_0^2=0.246$ GeV$^{2}$ to $\mu^2=16$ GeV$^{2}$ for the case of pion. The results are compared with the FNAL-E615 experimental data \cite{dy-exp4}, modified FNAL-E615 data \cite{Aicher:2010cb} and PDF obtained from BLFQ \cite{blfq}.}
\label{qcd-evolution}
\end{figure}

The valence quark distribution functions for the light pseudoscalar mesons can be computed using the parameters: constituent quark masses $(m, m_1, m_2)$ and the harmonic scale $(\beta_\pi, \beta_K)$. At fixed light-front time, the PDF gives the probablity of finding the quark in meson where the quark carries a longitudinal momentum fraction $x=k^+/P^+$. It is defined as \cite{pdf-def}
\begin{eqnarray}
f^{\mathcal{P}}(x)&=&\frac{1}{2}\int \frac{dz^-}{4 \pi} e^{ik^+ z^-/2} \times \bra{\mathcal{P}^+(P);S}\bar{\Psi}(0)\Gamma \Psi(z^-)\ket{\mathcal{P}^+(P);S}\vert_{z^+=\textbf{z}_\perp=0}.
\end{eqnarray}
The valence quark distribution for the case of pion and kaon are shown in Fig. \ref{qcd-evolution}(a). The model scale pion PDF is then evolved to the higher scale by using the NNLO DGLAP equations of QCD \cite{Dokshitzer:1977sg,Gribov:1972ri,Altarelli:1977zs} and is shown in Fig. \ref{qcd-evolution}(b). We compare our results with FNAL-E615 and FNAL-E615 modified experimental data \cite{dy-exp4, Aicher:2010cb}, also with the predictions using basis light-front quantization approach (BLFQ) \cite{blfq}.


\section{ Generalized Parton Distributions}
We have calculated one chirally-even $H(x,\zeta=0,t)$ and the other chirally-odd $E_{T}(x,\zeta=0,t)$ GPDs of valence quark for pion and kaon in LCQM by  restricting ourself to the DGLAP region i.e. $\zeta<x<1$.
The correlation to evaluate the chirally-even GPD is defined via the bilocal operator of light-front correlation functions of the vector current whereas the chirally-odd GPD is related to the tensor current.
We can derive the valence quark GPDs in terms of the overlap form of LCWFs. The quark polarization is taken along $y$-direction.
	
\begin{figure}
\centering
	\begin{minipage}[c]{1\textwidth}
		(a)\includegraphics[width=.38\textwidth]{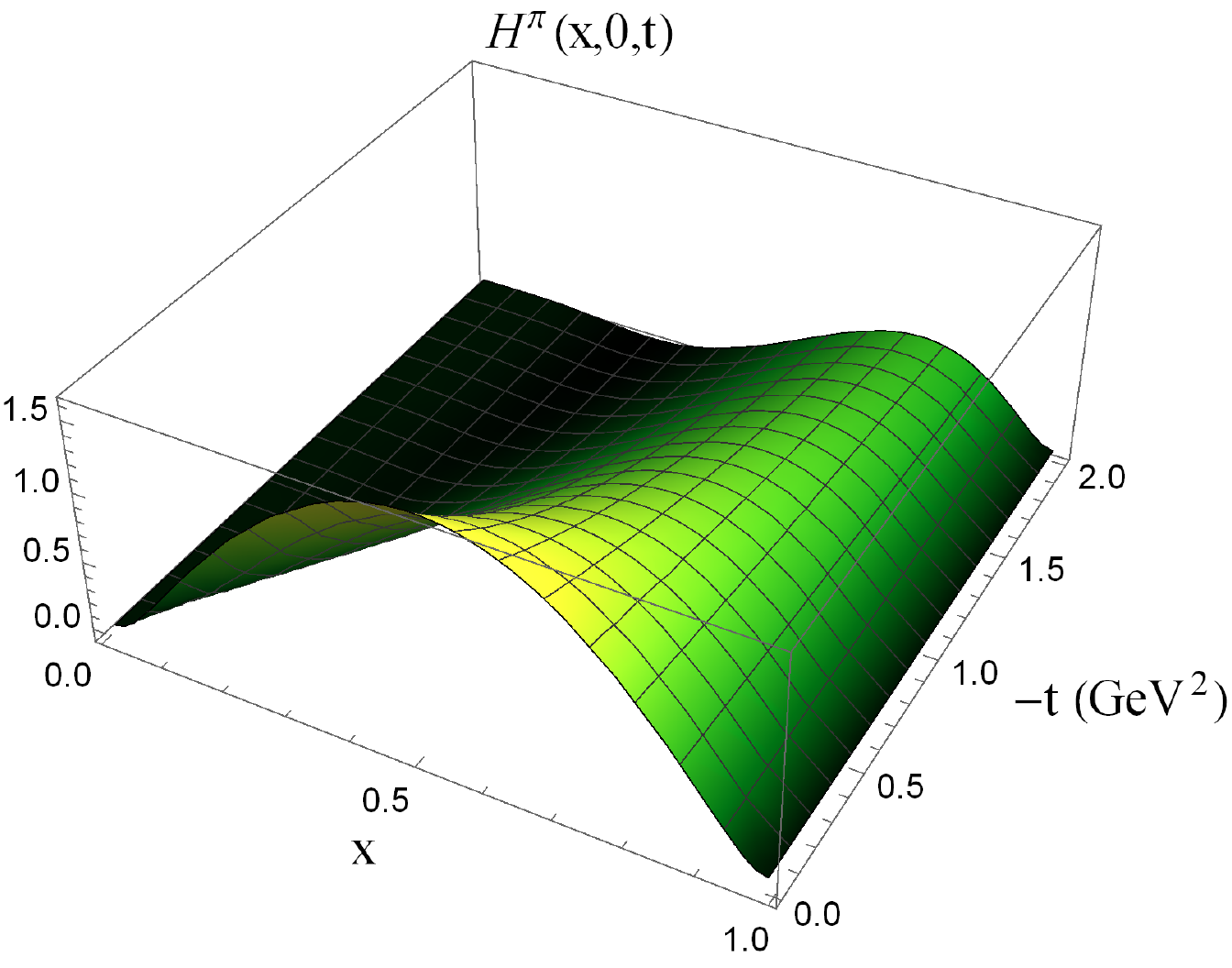}
		(b)\includegraphics[width=.38\textwidth]{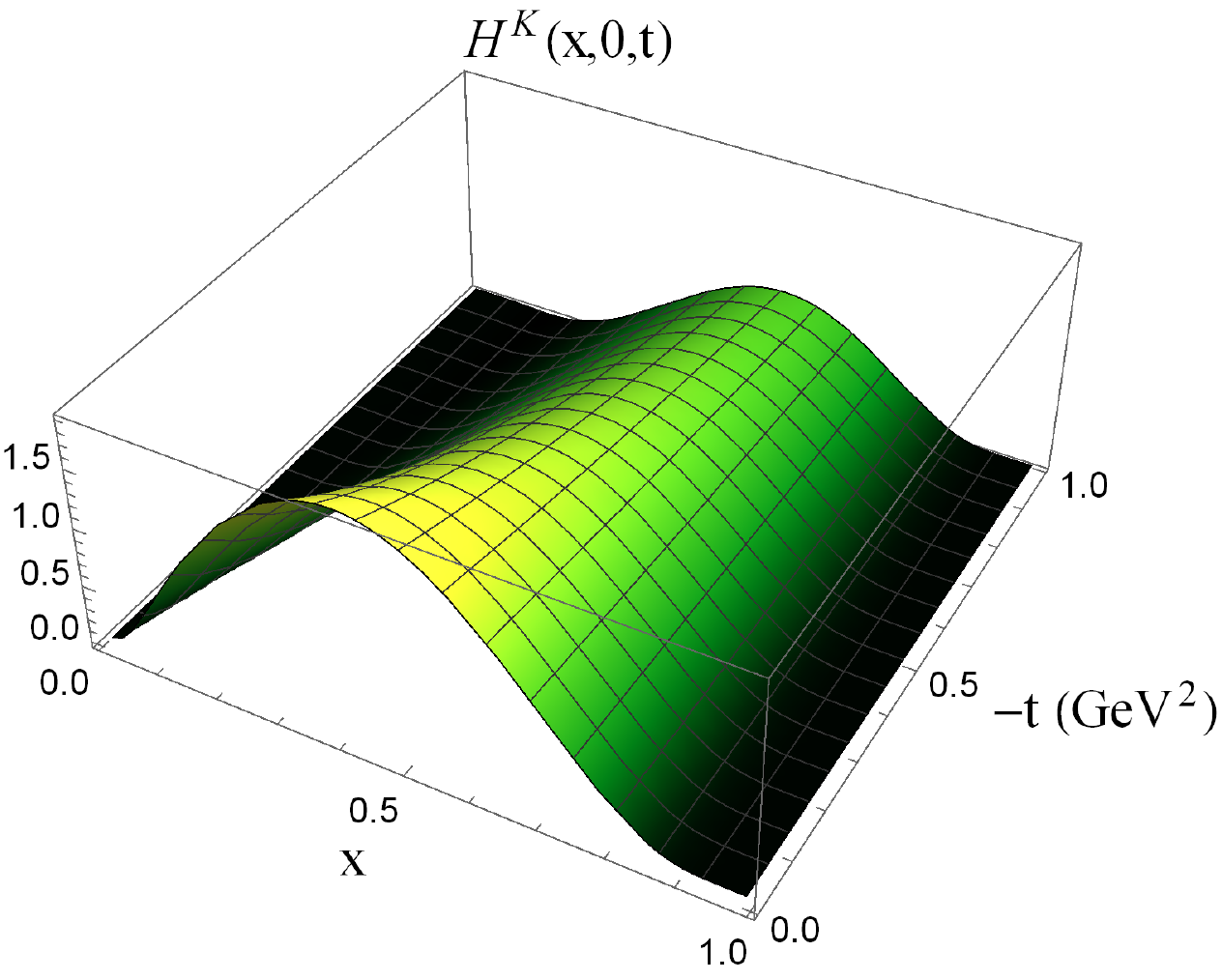}
		\end{minipage}
		\begin{minipage}[c]{1\textwidth}
		(c)\includegraphics[width=.38\textwidth]{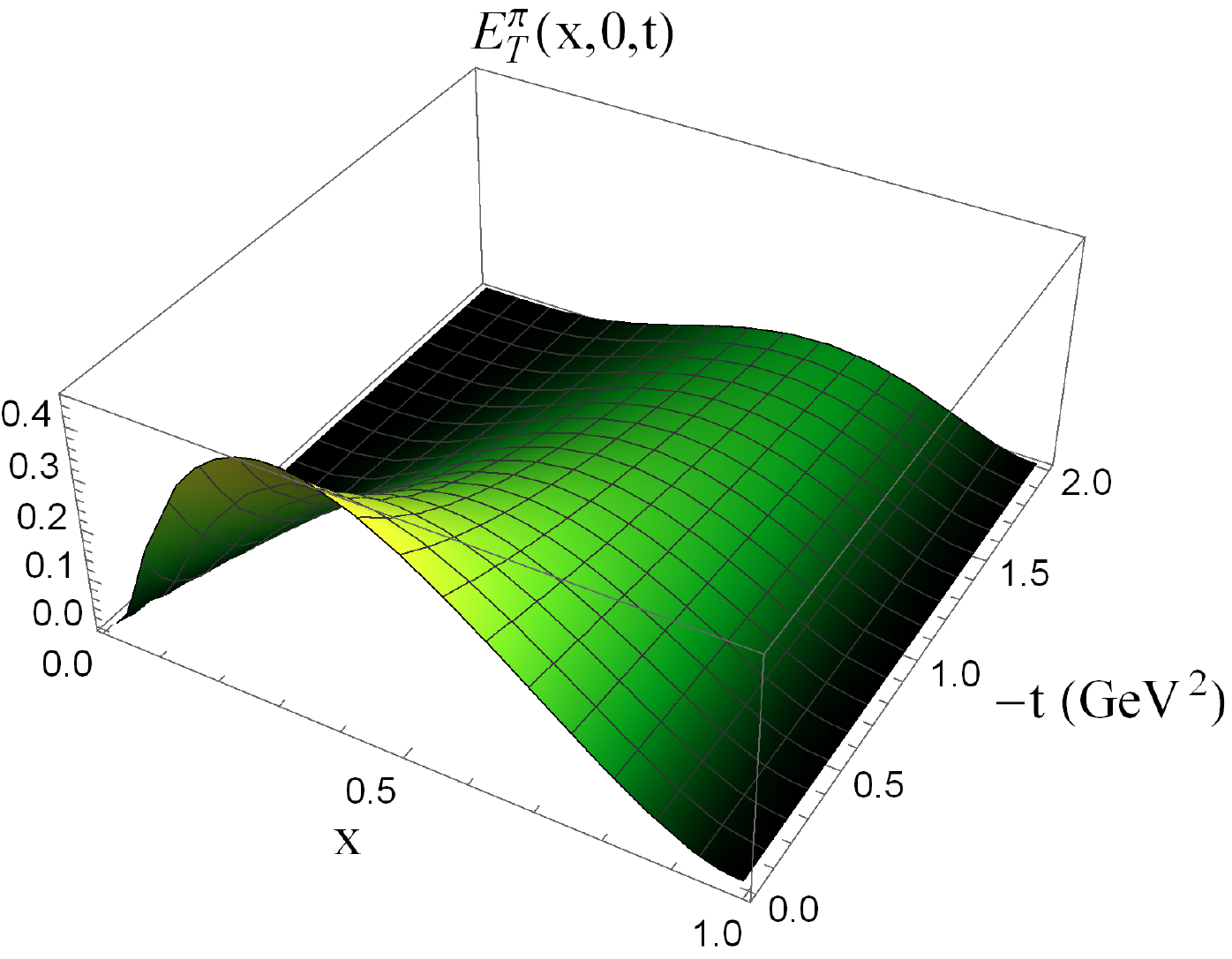}
		(d)\includegraphics[width=.38\textwidth]{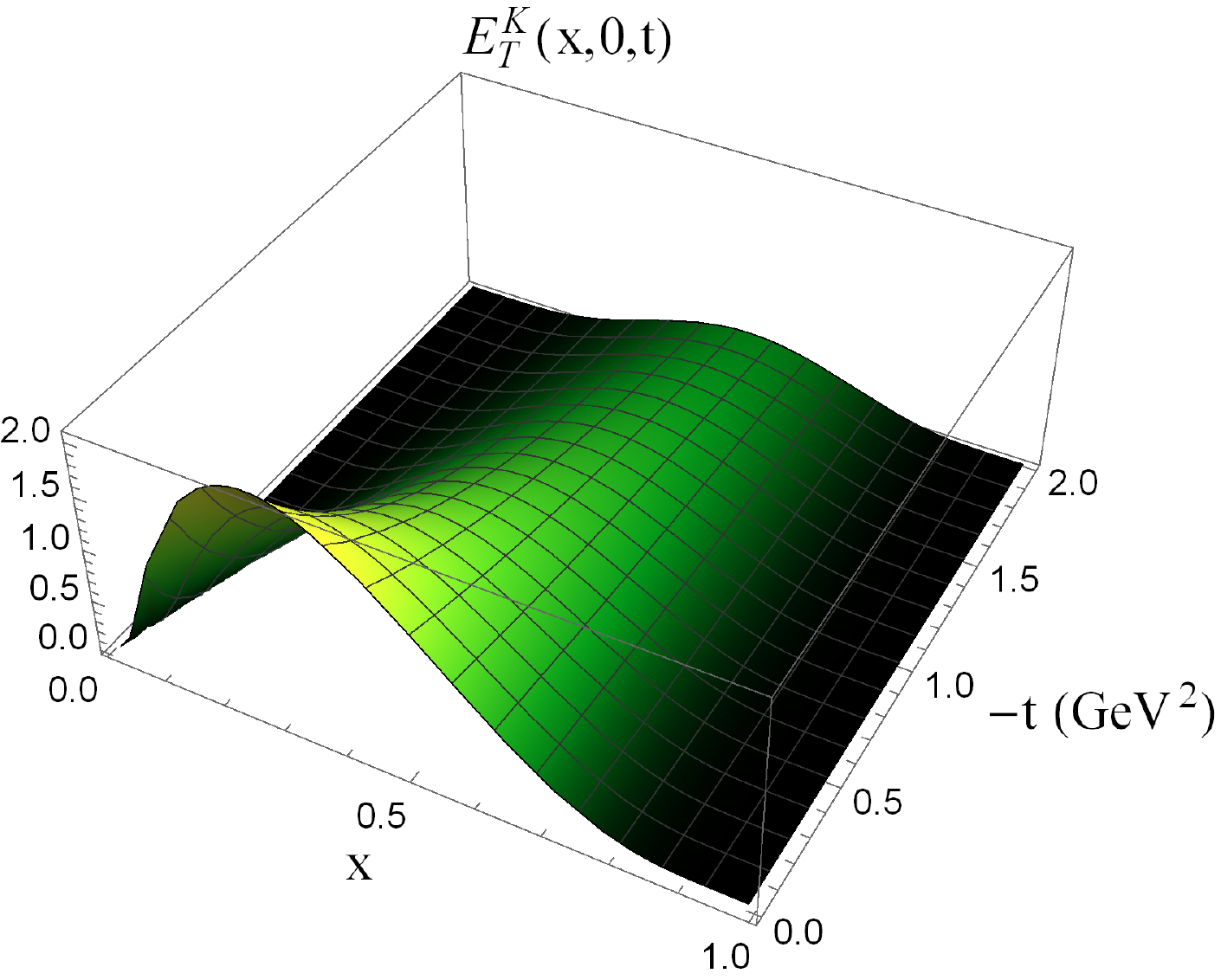}
	\end{minipage}
	\caption{The chiral-even and chiral-odd GPDs $H(x,0,t)$ and $E_T(x,0,t)$ with respect to $x$ and $-t$ (in $\rm GeV^2$) for pion (left panel) and kaon (right panel).}
	\label{gpd}
\end{figure}

In Fig. \ref{gpd}, we present the GPDs $H$ and $E_T$  for the case of pion (left panel) and kaon (right panel) with respect to $x$ and $-t$. The unpolarized quark distribution in the pion is maximum at the center of longitudinal momentum fraction ($x=0.5$) when the momentum transfer is zero. However, the peak of the chiral-odd GPD $E_T$ in pion appears when $x<0.5$. With increase in the momentum transfer $-t$, the magnitude of distribution peak becomes lower and shifts towards higher values of $x$.

\section{Conclusion}
We have presented various valence quark distributions suitable for low-resolution properties in the pion and the kaon using the light-cone quark model.
The results of pion PDF have been found to be in agreement with the reanalyzed E615 experimental data \cite{Aicher:2010cb}. We have evaluated GPDs in DGLAP region for zero skewness ($\zeta=0$) i.e. $0<x<1$, which provide 3D structure of mesons.
At zero momentum transfer, the chirally-even GPD lead to unpolarized quark distribution, $H(x, 0, 0) = f_1(x)$.
Depending upon the total momentum transferred to the pseudoscalar meson, we observe the change in distribution with respect to active quark longitudinal momentum fraction.

\section*{Acknowledgements}
The work of HD is supported by the Science and Engineering Research Board, Government of India, under MATRICS (Ref No. MTR/2019/000003). CM thanks the Chinese Academy of Sciences President's International Fellowship Initiative for the support via Grants No. 2021PM0023.

\end{document}